\documentclass[onecolumn,floatfix,preprintnumbers,eqsecnum,letterpaper,superscriptaddress,nofootinbib]{revtex4}
\usepackage{graphicx}
\usepackage{microtype}
\usepackage{amsmath}
\usepackage{amssymb}
\usepackage{subfigure}
\usepackage{hyperref}
\usepackage{url}
\usepackage{xcolor}
\usepackage{color}
\usepackage{mathrsfs}
\usepackage{calrsfs}
\usepackage{amsfonts}
\usepackage{eufrak}
\usepackage{tabularx}
\usepackage{eucal}
\usepackage{latexsym}
\usepackage{ragged2e}
\usepackage{epsfig}
\usepackage{textcomp}
\usepackage{caption}
\usepackage{subfigure}
\usepackage{marvosym}
\usepackage{ifsym}
\usepackage{lipsum}
\usepackage{appendix}
\usepackage{bm}
\usepackage{mathrsfs}

\DeclareCaptionJustification{justified}{\leftskip=0pt \rightskip=0pt \parfillskip=0pt plus 1fil}
\definecolor{vividviolet}{rgb}{0.62, 0.0, 1.0}
\definecolor{amaranth}{rgb}{0.9, 0.17, 0.31}
\definecolor{palatinateblue}{rgb}{0.15, 0.23, 0.89}
\definecolor{brightpink}{rgb}{1.0, 0.0, 0.5}
\definecolor{cornflowerblue}{rgb}{0.39, 0.58, 0.93}
\definecolor{deepcarminepink}{rgb}{0.94, 0.19, 0.22}
\definecolor{radicalred}{rgb}{1.0, 0.21, 0.37}

\hypersetup{ linktoc=all,
    colorlinks, linkcolor={palatinateblue},
    citecolor={brightpink}, urlcolor={amaranth}
}

\graphicspath{{Images/}}

\graphicspath{{Images/}}

\def\@fnsymbol#1{\ensuremath{\ifcase#1\or \ddagger \or  $\textleaf$  \or \dagger
\else\@ctrerr\fi}}%

\makeatother

\def\sideremark#1{\ifvmode\leavevmode\fi\vadjust{\vbox to0pt{\vss
 \hbox to 0pt{\hskip\hsize\hskip1em
 \vbox{\hsize1.3cm\tiny\raggedright\pretolerance10000
 \noindent #1\hfill}\hss}\vbox to8pt{\vfil}\vss}}}%

\def\beq{\begin{equation}}
\def\eeq{\end{equation}}

\begin{document}

\title{Cooling-Heating Properties of the FRW Universe in Gravity with a Generalized Conformal
Scalar Field}

\author{Haximjan Abdusattar \Letter}
\email{axim@nuaa.edu.cn}
\affiliation{College of Physics and Electrical Engineering, Kashi University, Kashi 844006, Xinjiang, China}

\author{Shi-Bei Kong}
\email{shibeikong@ecut.edu.cn}
\affiliation{School of Science, East China University of Technology, Nanchang 330013, Jiangxi, China}

{\let\thefootnote\relax\footnotetext{\vspace*{0.2cm}$^{\text{\Letter}}$ Corresponding Author}}

\begin{abstract}

In this paper, within the framework of modified gravity involving a conformal scalar field, we investigate the Joule-Thomson expansion of the FRW universe to identify cooling and heating regions. Notably, we observe that the Joule-Thomson coefficient, denoted as $\mu$, diverges at $R_A=\sqrt{-2\alpha}$ when $\alpha<0$, aligning with the thermodynamic singularity of the FRW universe. Additionally, we determine the inversion temperature and inversion pressure for the FRW universe, and illustrate the characteristics of inversion curves and isenthalpic curves in the $T$-$P$ plane. We compare these findings with results obtained under Einstein gravity, discussing the influence of the modification term on the cooling and heating properties of the FRW universe. This work contributes significantly to a deeper understanding of the formation of cooling and heating regions within the FRW universe, thereby advancing our comprehension of the physical mechanisms that govern the expansion of our universe.

\end{abstract}

\maketitle

\section{Introduction}

The Friedmann-Robertson-Walker (FRW) universe is a dynamical and spherically symmetric spacetime with an apparent horizon \cite{Bak:1999hd,Cai:2005ra}. It exhibits thermodynamic properties such as thermal radiation, Hawking temperature \cite{Cai:2008gw,Zhu:2009wa}, entropy, quasi-local Misner-Sharp energy, the unified first law, and work density \cite{Cai:2006rs}, etc. The way in which the first law of thermodynamics is obtained at the apparent horizon of the FRW universe is similar to those of black holes \cite{Hayward:1993wb,Hayward:1997jp}. Additionally, from the first law of thermodynamics and the Clausius relation, one can derive the Friedmann's equations in Einstein gravity and some modified theories of gravity \cite{Akbar:2006kj,Akbar:2006mq,Akbar:2006er,Gong:2007md}. Exploring the thermodynamics of the FRW universe uncovers an inherent link between thermodynamics and gravity, potentially constituting a fundamental and universal revelation. As a complete and self-consistent theory of quantum gravity remains elusive, delving deeper into the thermodynamics of FRW spacetime can shed light on certain aspects of quantum gravity effects. This research holds significant relevance for physicists in their quest to establish a cohesive quantum gravity framework.

Nevertheless, in addition to the thermodynamic laws of the FRW universe, various other intriguing phenomena, such as the thermodynamic equation of state, as well as $P$-$V$ phase transitions
\cite{Kubiznak:2012wp,Cai:2013qga,Xu:2015rfa,Cai:2014znn,Kubiznak:2016qmn,Kubiznak:2014zwa,Gunasekaran:2012dq,Wei:2012ui,Cheng:2016bpx,Wei:2020poh,Hendi:2012um,Hendi:2017fxp,Altamirano:2013ane,Altamirano:2014tva,Bhattacharya:2017nru,Li:2020xkh,Hu:2020pmr,Hu:2018qsy,Abdusattar:2023xxs} and Joule-Thomson expansion \cite{Okcu:2016tgt,Okcu:2017qgo,Lan:2018nnp,Pu:2019bxf,Li:2019jcd,Rajani:2020mdw,Bi:2020vcg,Ghaffarnejad:2018exz,Chabab:2018zix,MahdavianYekta:2019dwf,Mo:2018rgq,Kuang:2018goo,Xing:2021gpn,Cisterna:2018jqg,Liang:2021xny}, which have been observed in black holes, remain rarely explored within the context of the FRW universe. To characterize the behavior of a thermodynamic system fully, it is essential to establish its thermodynamic equation of state, alongside adhering to thermodynamic laws. However, in the case of the FRW universe, this proves challenging unless an appropriate definition of thermodynamic pressure can be determined. The definition of pressure utilized in the extended phase space thermodynamics of AdS black holes, i.e. the cosmological constant \cite{Kastor:2009wy,Dolan:2010ha,Dolan:2011xt,Cvetic:2010jb}, may not be directly applicable to the FRW spacetime, especially when the cosmological constant is set to zero.

In light of this challenge, we conducted a fundamental investigation in our recent paper \cite{Abdusattar:2021wfv}, addressing the thermodynamic pressure and equation of state for the FRW universe within the framework of Einstein's gravity, while considering a perfect fluid.\footnote{See the more related work in the literature \cite{Abdusattar:2022bpg}.}
Specifically, we delved into the Joule-Thomson expansion of the FRW universe to examine its cooling and heating properties. Our investigation led us to determine the inversion temperature, revealing that the transition between cooling and heating in the FRW universe corresponds to the phase transition between deceleration and acceleration. Encouraged by these findings, we further study the Joule-Thomson expansion of the FRW universe in the brane world scenario, and find the existence of an inversion point that is affected by the brane tension \cite{Kong:2022xny}, which may provide new ways to test the brane world scenario\footnote{See the related works in the literature \cite{Abdusattar:2015azp} and the references therein.}.
Besides, we also extended our exploration to the thermodynamics of the FRW universe within modified gravity theories, leading to the discovery of $P$-$V$ phase transitions in our recent works \cite{Kong:2021dqd,Abdusattar:2023pck,Abdusattar:2023hlj}. The subsequent series of studies can be found in literature
\cite{Saavedra:2023lds,Housset:2023jcm}.

In this paper, we present the Joule-Thomson expansion for the FRW universe within the framework of modified gravity involving a generalized conformal scalar field from Horndeski class \cite{Fernandes:2021dsb}. Notably, Horndeski gravity is an extensive scalar-tensor theory that permits the inclusion of high-order derivatives in its action. However, it manages to avoid the problematic Ostrogradsky instabilities by ensuring that its equations of motion involve, at most, second-order derivatives, as discussed in \cite{Kobayashi:2019hrl}.
This arouses our interest in determining whether the Joule-Thomson expansion as well as inversion temperature and inversion pressure can be obtained in the FRW universe under this modified gravity.
We carefully analyze the sign of Joule-Thomson coefficient $\mu$ to identify regions of cooling and heating. As the Joule-Thomson coefficient, $\mu$ offers insights into the temperature variations that occur during the expansion process. A thorough examination of this parameter enables a deeper understanding of the formation of cooling and heating regions within the FRW universe. This line of inquiry holds significant implications for our comprehension of the physical mechanisms that govern the expansion of our universe.

This paper is organized as follows. In Sec.\ref{SecII}, as a warmup exercise, we review briefly the essentials of the thermodynamics as well as equation of state for the FRW universe in modified gravity with a generalized conformal scalar field.
In Sec.\ref{SecIII}, we study the Joule-Thomson expansion of the FRW universe as an application of thermodynamic equation of state, and discuss the inversion temperature and inversion pressure. Sec.\ref{SecIV} is devoted to our conclusions and discussions.

\section{Brief Review on Thermodynamics and Equation of State for the FRW Universe}\label{SecII}

In this section, we make a brief review on the essentials of thermodynamics and equation of state for the spatially flat FRW universe in gravity with a generalized conformal scalar field, in which the thermodynamic pressure $P$ defined by the work density $W$ of the perfect fluid, i.e., $P\equiv W$ \cite{Kong:2021dqd,Abdusattar:2023pck}.

\subsection{Brief Introduction of Gravity with a Generalized Conformal Scalar Field and Friedmann's Equation's}

We initiate our study by introducing a well defined action \cite{Fernandes:2021dsb}
\begin{eqnarray}\label{eq:actionconfgeneral}
S=\int \frac{d^{4} x \sqrt{-g}}{16\pi}\Big[\mathcal{R}-2\Lambda -\beta e^{2\phi}\left(\mathcal{R} + 6(\nabla \phi)^{2}\right)-2\lambda e^{4\phi}
- \alpha \Big(\phi \mathcal{G} - 4 G^{\mu \nu} \nabla_{\mu} \phi \nabla_{\nu} \phi - 4 \square \phi(\nabla \phi)^{2} - 2(\nabla \phi)^{4}\Big)\Big]+S_{m}\,
\end{eqnarray}
which defines a scalar-tensor theory that is characterized by $\alpha$, $\beta$, and $\lambda$. Here $\mathcal{G}\equiv\mathcal{R}^2-4 R^{\mu \nu} R_{\mu \nu}+R^{\mu \nu \alpha \beta} R_{\mu \nu \alpha \beta}$ is the Gauss-Bonnet term, $g$ is the determinant of metric tensor $g_{\mu\nu}$, $G^{\mu \nu}$ is the Einstein tensor, $\mathcal{R}$ is the Ricci scalar, and the action $S_{m}$ associated with the matter field. In this context, $\square$ represents the covariant d'Alembert operator, and $(\nabla \phi)^2$ is the short form of
$g^{\mu\nu}\nabla_{\mu}\phi\nabla_{\nu}\phi$, where $\nabla_\mu$ denotes the covariant derivative, and $\phi$ is represents a scalar field.

The field equations in this modified gravity are obtained as
\begin{equation}\label{FieldEq}
G_{\mu \nu} + \Lambda g_{\mu \nu} +\alpha \mathcal{H}_{\mu \nu} - \beta e^{2\phi} \mathcal{A}_{\mu \nu}+\lambda e^{4\phi}g_{\mu \nu}=8\pi T_{\mu\nu}\,,
\end{equation}
where $T_{\mu\nu}$ is the stress-energy tensor of matter, and
\begin{eqnarray}
\mathcal{H}_{\mu\nu} &=& 2G_{\mu \nu} (\nabla\phi)^2+4P_{\mu \alpha \nu \beta}(\nabla^\alpha \phi \nabla^\beta \phi - \nabla^\beta \nabla^\alpha \phi) +4(\nabla_\alpha \phi \nabla_\mu \phi - \nabla_\alpha \nabla_\mu \phi) (\nabla^\alpha \phi \nabla_\nu \phi - \nabla^\alpha \nabla_\nu \phi)\nonumber\\
&+&4(\nabla_\mu \phi \nabla_\nu \phi - \nabla_\nu \nabla_\mu \phi) \square\phi+g_{\mu \nu} [2(\square\phi)^2 - (\nabla \phi)^4] +g_{\mu \nu} [2\nabla_\beta \nabla_\alpha \phi (2\nabla^\alpha \phi \nabla^\beta \phi - \nabla^\beta \nabla^\alpha \phi)]\,,\nonumber
\end{eqnarray}
\begin{equation}
\mathcal{A}_{\mu\nu} = G_{\mu \nu} + 2\nabla_\mu \phi \nabla_\nu \phi - 2\nabla_\mu \nabla_\nu \phi +g_{\mu \nu} [2\square\phi + (\nabla\phi)^2] \,,\nonumber
\end{equation}
with
\begin{eqnarray}
P_{\alpha \beta \mu \nu} &\equiv& *R*_{\alpha \beta \mu \nu} = -R_{\alpha \beta \mu \nu}-g_{\alpha \nu} R_{\beta \mu}+g_{\alpha \mu} R_{\beta \nu} \nonumber\\
&&-g_{\beta \mu} R_{\alpha \nu}+g_{\beta \nu} R_{\alpha \mu}-\frac{1}{2}\left(g_{\alpha \mu} g_{\beta \nu}-g_{\alpha \nu} g_{\beta \mu}\right) \mathcal{R}\,.\nonumber
\end{eqnarray}
The distinctive feature of the construction in \cite{Fernandes:2021dsb} lies in the combination of the trace of the metric equations and the scalar field equation derived from the action (\ref{eq:actionconfgeneral}), resulting in a four-dimensional equation that purely relies on geometric principles,
\begin{equation}\label{Eq:trace}
\mathcal{R}+\frac{\alpha}{2}\mathcal{G}-4\Lambda = -8\pi T \,,
\end{equation}
which closely resembles the trace equation found in the higher-dimensional Einstein-Gauss-Bonnet theory, where $T=g^{\mu\nu}T_{\mu\nu}$ represents the trace of the stress-energy tensor.

In the co-moving coordinate system $\{t,r,\theta,\varphi\}$, the line element of the spatially flat ($k=0$) \footnote{Note that the values $k=-1, +1, 0$ represent the spatial curvatures associated with open (Hyperbolic), closed (Spherical), and flat universes, respectively. In this paper, our sole focus is on the case where $k=0$.} FRW universe can be written as
\begin{equation}
d s^2=-d t^2+a^2(t)[d r^2+r^2(d\theta^2+\sin^2\theta d\varphi^2)]\,, \label{LE}
\end{equation}
where $a(t)$ is the time-dependent scale factor. We  consider the matter content of the FRW universe to be described by a perfect fluid with a stress-energy momentum tensor
\begin{equation}
T_{\mu\nu}=(\rho+p)u_{\mu}u_{\nu}+p g_{\mu\nu}\,,\label{ST}
\end{equation}
where $\rho$ and $p$ are energy density and pressure of the perfect fluid, $u_{\mu}$ is the four velocity.
For simplicity, we employ the stress-tensor (\ref{ST}) for the perfect fluid and incorporate the cosmological constant term into it. Consequently, in the subsequent discussion, the presence of $\Lambda$ is not explicitly evident. Under this scenario, the Friedmann's equations for the spatially flat FRW universe take the following form
\cite{Fernandes:2021dsb}
\begin{eqnarray}
(1+\alpha H^2)H^2&=&\frac{8\pi}{3}\rho\,, \label{GBrhom}\\
(1+2\alpha H^2)\dot{H}&=&-4\pi(\rho+p)\,, \label{GBpm}
\end{eqnarray}
where $H\equiv \dot{a}(t)/a(t)$ is the expansion rate of the universe.
Interestingly, these Friedmann's equations bear a resemblance to those found in holographic cosmology \cite{Apostolopoulos:2008ru,Bilic:2015uol}, quantum-corrected entropy-area relations \cite{Cai:2008ys}, generalized uncertainty principles \cite{Lidsey:2009xz}, and four-dimensional Einstein-Gauss-Bonnet gravity \cite{Feng:2020duo}. Consequently, the thermodynamics of the FRW universe in this model shares similar properties with these theories.

\subsection{Thermodynamics of FRW Universe in Gravity with a Generalized Conformal Scalar Field}

For convenience, we introduce another form of the line-element (\ref{LE}) with the areal radius defined as $R\equiv a(t)r$,
\begin{equation}\label{NewFRW}
d s^2=h_{ij}d x^i d x^j+R^2(d\theta^2+\sin^2\theta d\varphi^2)\,,
\end{equation}
where $i,j=0,1$ with $x^0=t, x^1=r$ and $h_{ij}=diag[-1,a^2(t)]$. By using the metric (\ref{NewFRW}), it becomes straightforward to determine the apparent horizon of the FRW universe through the solution of $h^{ij}\partial_i R\partial_j R=0$ \cite{Hayward:1993wb}. This for the line-element (\ref{NewFRW}) yields the following result \cite{Bak:1999hd,Cai:2005ra}
\begin{equation}
R_A=\frac{1}{H}\,, \label{AH}
\end{equation}
whose time derivative is then
\begin{equation}
\dot{R}_A=-\dot{H}R^2_A \,. \label{dot}
\end{equation}

For the spherically symmetric spacetime, the work density of the matter defined as $W:=-\frac{1}{2}h_{ij}T^{ij}$ \cite{Hayward:1997jp}. This for the FRW universe can be easily obtained by using the Eqs.(\ref{GBrhom}), (\ref{GBpm}), (\ref{AH}) and (\ref{dot}) \cite{Kong:2021dqd,Abdusattar:2023pck}
\begin{eqnarray}\label{WD}
W=\frac{1}{2}(\rho-p)
=\frac{3}{8\pi R^2_A}+\frac{3\alpha}{8\pi R^4_A}-\frac{\dot{R}_A}{8\pi R^2_A} \Big(1+\frac{2\alpha}{R^2_A}\Big)\,,
\end{eqnarray}
which will serve as the basis of  establishing the definition of thermodynamic pressure in the FRW universe. Here, $T^{ij}$ is the projection of the stress energy-momentum tensor $T_{\mu \nu}$ in the $(t,r)$ direction.

For a dynamical and spherically symmetric spacetime, its surface gravity is defined as $\kappa={\partial_i}(\sqrt{-h}~h^{ij}{\partial_{j}R})/{(2\sqrt{-h})}$ \cite{Hayward:1997jp},
where $h={\rm det}(h_{ij})$. By utilizing (\ref{NewFRW}), (\ref{AH}), and (\ref{dot}), the surface gravity of the spatially flat FRW universe at the apparent horizon can be derived \cite{Cai:2005ra,Cai:2006rs,Akbar:2006kj}
\begin{equation}\label{SurfaceG}
\kappa|_{R=R_{A}}=-\frac{1}{R_A}\Big(1-\frac{\dot{R}_A}{2}\Big)\,.
\end{equation}
Assuming $\dot{R}_A$ to be a small quantity, the surface gravity $\kappa$ of the apparent horizon in the FRW universe is negative \cite{Hayward:1993wb,Abdusattar:2021wfv,Abdusattar:2022bpg,Dolan:2013ft}.
It is straightforward to deduce that the Hawking temperature of the spatially flat FRW universe at the apparent horizon in the following
\begin{equation}
T\equiv -\frac{\kappa|_{R=R_{A}}}{2\pi}=\frac{1}{2\pi R_A}\Big(1-\frac{\dot{R}_A}{2}\Big)\,. \label{HT}
\end{equation}

From (\ref{GBrhom}) with (\ref{AH}) and thermodynamic volume $V\equiv {4}\pi R_A^3/3$, the energy for the FRW universe obtained by
\begin{eqnarray}\label{E}
E=\rho V=\frac{R_A}{2}+\frac{\alpha}{2R_A}\,,
\end{eqnarray}
which could be regarded as the effective Misner-Sharp energy \cite{Maeda:2007uu,Cai:2009qf}, and satisfy the first law
\begin{equation}
d E=-T d S+W d V\,, \label{FL}
\end{equation}
where the entropy is given by
\begin{eqnarray}\label{SV}
S= \frac{A}{4}+2\pi \alpha \ln \Big(\frac{A}{A_0}\Big)\,
\end{eqnarray}
with $A = 4\pi R_A^2$ being the area of the apparent horizon and $A_0$ can take any positive constant with the dimensionality of area to guarantee the logarithmic function is well defined.
Note that the form of entropy is the same as the black hole entropy with quantum correction
\cite{Cai:2008ys,Solodukhin:1997yy,Mann:1997hm,Das:2001ic,Mukherji:2002de,Gour:2003jj,Chatterjee:2003uv} and the entropy of four dimensional Gauss-Bonnet black hole in AdS space \cite{Wei:2020poh}. One important aspect to note is that the negative sign preceding $T dS$ in equation (\ref{FL}) is a result of negative surface gravity, i.e. $\kappa<0$, as we previously discussed below Eq.(\ref{SurfaceG}). Therefore, in order to ensure a positive temperature $T$, this minus sign is retained, as explained in references \cite{Banihashemi:2022htw,Abdusattar:2021wfv,Abdusattar:2022bpg}.

Comparing the Eq.(\ref{FL}) with the standard form of thermodynamic first law
\begin{eqnarray}\label{FL}
d U=T d S-P d V\,,
\end{eqnarray}
one can see that the energy $E$ should be interpreted as the minus of the internal energy $U:=-E$, and the work density $W$ as the thermodynamic pressure $P$, i.e., $P:=W$.
After establishing the definition of thermodynamic pressure, with the help of Eqs.(\ref{WD}) and (\ref{HT}), one can obtain the equation of state for the FRW universe in modified gravity with a conformally scalar field given by \footnote{If $\alpha<0$, the equation of state for $R_A=\sqrt{-2\alpha}$ is independent of temperature, refering to a thermodynamic singularity.}\cite{Kong:2021dqd,Abdusattar:2023pck}
\begin{eqnarray}\label{EoS}
P=\frac{T}{2R_A}\Big(1+\frac{2\alpha}{R_A^2}\Big)+\frac{1}{8\pi R^2_A}\Big(1-\frac{\alpha}{R^2_A}\Big)\,,\label{EoS}
\end{eqnarray}
which recovers to the results in Einstein's theory of gravity \cite{Abdusattar:2021wfv} when $\alpha=0$. In the next section, we explore the effects of the gravitational coupling term on thermodynamic properties of the FRW universe.

\section{Joule-Thomson Expansion of the FRW Universe in Gravity with a Generalized Conformal Scalar Field}\label{SecIII}

In this section, we study the cooling and heating properties of the FRW universe through the Joule-Thomson expansion.
A noteworthy similarity exists between the thermodynamic equation of state for the FRW universe and that of a van der Waals system.
An interesting feature of a van der Waals system is the occurrence of either heating or cooling during the throttling process.
It is worth noting that the throttling process of van der Waals system is already applied to AdS black holes \cite{Okcu:2016tgt,Okcu:2017qgo,Lan:2018nnp,Pu:2019bxf,Li:2019jcd,Rajani:2020mdw,Bi:2020vcg,Ghaffarnejad:2018exz,Chabab:2018zix,MahdavianYekta:2019dwf,Mo:2018rgq,Kuang:2018goo,Xing:2021gpn,Cisterna:2018jqg,Liang:2021xny}.
Therefore, it is valuable to examine the Joule-Thomson expansion of the FRW universe in the context of modified gravity.

The notable aspect of this adiabatic, irreversible expansion is that the enthalpy remains unchanged in both the initial and final states. As a result, the isenthalpic curves, representing all points associated with the initial and final equilibrium states at the same enthalpy, are essentially constant enthalpy curves. This expansion is accompanied by a change in temperature, thus defining the Joule-Thomson coefficient \cite{Okcu:2016tgt}
\begin{equation}\label{JTC}
\mu:=\large \left(\frac{\partial T}{\partial P}\large \right)_{\cal H} \,,
\end{equation}
which distinguishes between the cooling and heating behaviors of a thermal system. The inversion temperature, denoted as $T_i$, marks the temperature at which the Joule-Thomson coefficient becomes zero, i.e., $\mu(T_i)=0$. The behavior can be divided into two distinct regimes based on the value of $T_i$: when $T<T_i$ ($T>T_i$), the Joule-Thomson process induces cooling (heating) of the system, characterized by $\Delta T<0$ and $\mu>0$ ($\Delta T>0$ and $\mu<0$), respectively. As the system's temperature converges toward $T_i$, the pressure corresponds to the inversion pressure $P_i$, establishing the inversion point $(T_i,P_i)$ where the transition from cooling to heating takes place.

Use Eqs.(\ref{E}), (\ref{EoS}), we obtain the enthalpy of FRW universe given by
\begin{eqnarray}\label{HH}
 {\cal H}&\equiv&U+PV\nonumber\\
 &=& \frac{(2\pi R_{A} T-1)(R_A^2+2\alpha)}{3R_A}\,.
\end{eqnarray}
When $\alpha>0$, the enthalpy is positive ${\cal H}>0$ for $R_A T>1/2\pi$, and negative ${\cal H}<0$ for $0<R_A T<1/2\pi$. \\
When $\alpha<0$,
 \begin{itemize}
  \item the enthalpy ${\cal H}>0$ if $R_A>\sqrt{-2\alpha}$, and negative ${\cal H}<0$ if $R_A<\sqrt{-2\alpha}$, for $R_A T>1/2\pi$.
  \item the enthalpy ${\cal H}>0$ if $R_A<\sqrt{-2\alpha}$, and negative ${\cal H}<0$ if $R_A>\sqrt{-2\alpha}$, for $0<R_A T<1/2\pi$.
  \item the enthalpy ${\cal H}=0$ if $R_A=\sqrt{-2\alpha}$.
\end{itemize}

From (\ref{HH}), the temperature is a function of ${\cal H}$ and $R_A$ as follows
\begin{equation} \label{THRA}
T({\cal H},R_A)=\frac{R_A^2+3{\cal H}+2\alpha}{2\pi R_A(R_A^2+2\alpha)}\,.
\end{equation}
Then substituting (\ref{PHRA}) into the (\ref{EoS}), the pressure $P$ can be rewritten as a function of ${\cal H}$ and $R_A$,
\begin{equation} \label{PHRA}
P({\cal H},R_A)=\frac{3(R_A^2+2{\cal H}R_A+\alpha)}{8\pi R_A^4} \,.
\end{equation}
By using (\ref{JTC}) combined with Eqs.(\ref{THRA}) and (\ref{PHRA}), we obtain the Joule-Thomson coefficient of the FRW universe in this modified gravity
\begin{eqnarray}\label{JTFRW}
\mu&=&\large \frac{\left({\partial T}/{\partial R_A}\large \right)_{{\cal H}}}{\large \left({\partial P}/{\partial R_A}\large \right)_{{\cal H}}}\nonumber\\
&=&\frac{2R_A^3}{R_A^2+3{\cal H}R_A+2\alpha}\left[1+\frac{6{\cal H}R_A^3}{(R_A^2 +2\alpha)^2}\right] \nonumber\\
&=&\frac{R_A^4(4\pi R_A T-1)+2R_A^2 \alpha}{3\pi T(R_A^2+2\alpha)^2} \, .
\end{eqnarray}
In particular, the divergence point of Joule-Thomson coefficient is consistent with the thermodynamic singular point at $R_A=\sqrt{-2\alpha}$ for $\alpha<0$.
Except for the singular point, setting $\mu=0$, we can obtain the enthalpy
\begin{equation}
{\cal H}=-\frac{(R_A^2+2\alpha)^2}{6R_A^3}<0 \,,
\end{equation}
which indicate that the inversion point exists just for the negative enthalpy. The inversion point corresponds to $\ddot{a}(t)=-2\alpha \dot{a}^4(t)/a^3(t)$, which derivations and physical interpretations are discussed in the Appendix \ref{appA}.
Thus, we easily find that the inversion temperature and inversion pressure both exist if the enthalpy ${\cal H}$ for a FRW universe is negative.
In this situation, from (\ref{JTFRW}), we get the inversion temperature
\begin{eqnarray}\label{invT}
T_{i}= \frac{R_A^2 -2\alpha}{4\pi R_A^3}\,,
\end{eqnarray}
and the inversion pressure
\begin{equation}\label{invP}
P_{i}= \frac{2R_A^4 -R_A^2 \alpha -4\alpha^2}{8\pi R_A^6}\,,
\end{equation}
where we have used Eq.(\ref{EoS}).

\begin{figure*}[ht]
\centering
 \begin{minipage}[t]{1\linewidth}
\includegraphics[width=8cm]{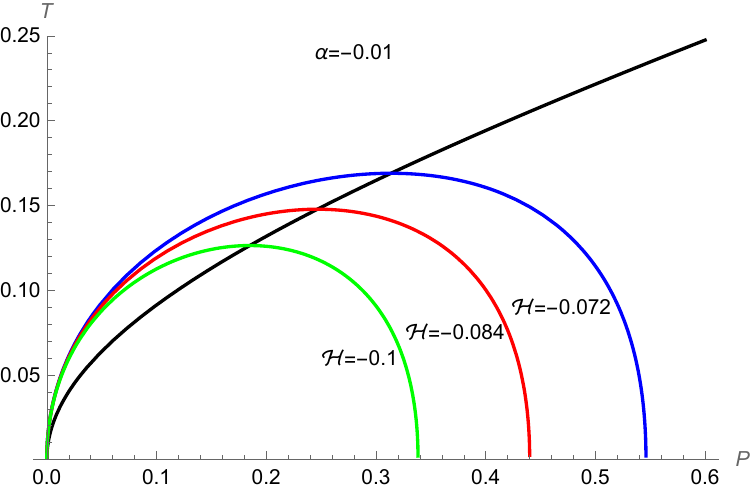}
 \put(-128,-10){(a)}
 \includegraphics[width=8cm]{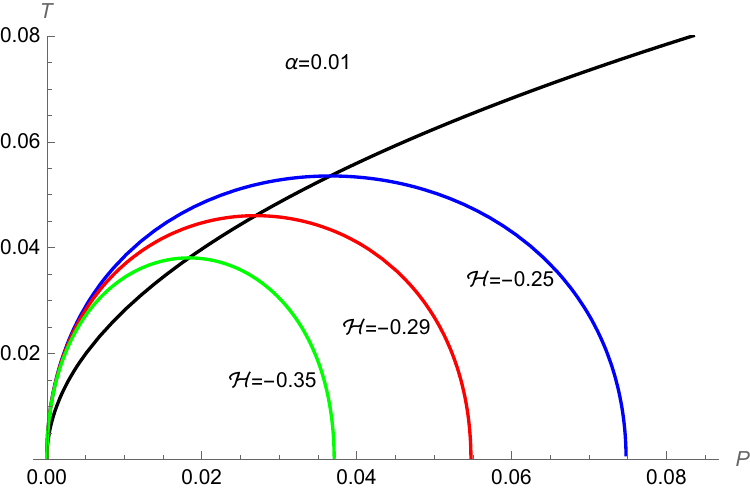}
 \put(-128,-10){(b)}
\end{minipage}
\centering
 \begin{minipage}[t]{1\linewidth}
\includegraphics[width=8cm]{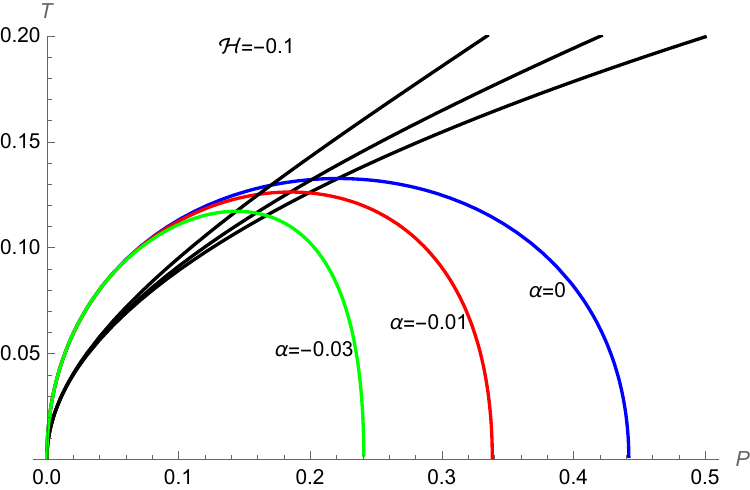}
\put(-128,-10){(c)}
\includegraphics[width=8cm]{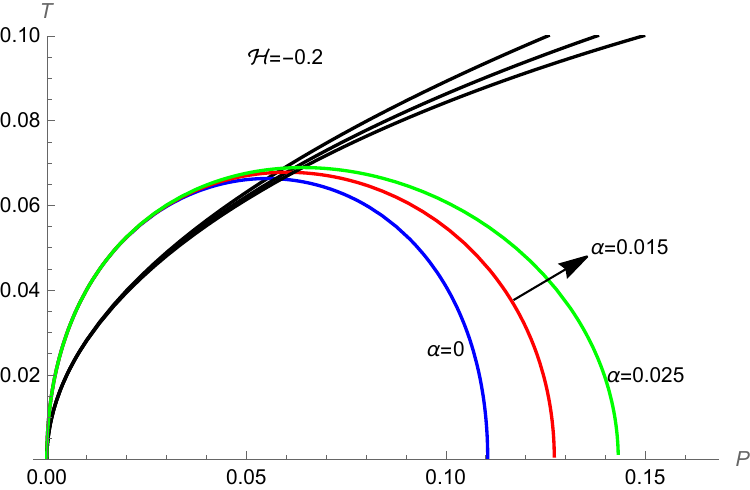}
\put(-128,-10){(d)}
\end{minipage}
 \caption{\label{FigJT} \footnotesize The inversion and isenthalpic curves of the FRW universe with different values of ${\cal H}$ and $\alpha$. The black lines are the inversion curves. The rainbow curves intersects with the inversion curves, at which point it has maxima. The values of ${\cal H}$ and $\alpha$ are marked on $T$-$P$ plane.}
\end{figure*}

Considering the Joule-Thomson expansion is an isenthalpic process, it is intriguing to investigate the isenthalpic curves of the FRW universe. In the followings, we plot the inversion and isenthalpic curves of the FRW universe in $T$-$P$ plane with fixed ${\cal H}$, and determine the cooling-heating region. Figure \ref{FigJT} shows the inversion and isenthalpic curves, which are plotted by Eqs. (\ref{THRA}),(\ref{PHRA}) and (\ref{invT}), (\ref{invP}).

The black line represents the inversion curve, the rainbow curves represents isenthalpic curves.
We can see that the inversion curve intersects with the maximum points of isenthalpic curves, and it would separate the plane into the cooling and heating regions. At that point the Joule-Thomson coefficient is zero, which indicates that the cooling-heating transition coincides at the maximum points of isenthalpic curves.
In fact, the inversion curve acts as a boundary between the cooling and heating regions, and cooling (heating) does not occur on the inversion curve. Therefore, we can distinguish between the cooling and heating region by checking the sign of the slope of the isenthalpic curves. The positive sign of slope stands for the cooling region and the minus for the heating region. We can  see that the inversion temperature $T_i$ increase as the $\alpha$ and ${\cal H}$ increase, at the lower pressure area.
On the other hand, we see from Fig.\ref{FigJT}(a)(b) that the region surrounded by isenthalpic curve and $P$ axis shrinks when ${\cal H}$ decreases.
Moreover, we can see from Fig.\ref{FigJT}(c)(d) that the values of the inversion temperature and inversion pressure get larger with the increasing of $\alpha$. What's more,
if $\alpha$ is negative, the values of the inversion points are smaller than those in Einstein gravity (i.e.,
$\alpha=0$). Conversely, if $\alpha$ is positive, the values of the inversion points are larger than those in Einstein gravity (i.e., $\alpha=0$).

\section{Conclusions and Discussions} \label{SecIV}

In this paper, by considering work density as thermodynamic pressure, we investigated the Joule-Thomson expansion of the FRW universe with a perfect fluid in modified gravity with generalized conformal scalar field. Firstly, according to the thermodynamic quantities of the FRW universe, we obtain the Joule-Thomson coefficient $\mu$. We find that the coefficient $\mu$ has a divergence point at $R_A=\sqrt{-2\alpha}$ for $\alpha<0$, which is consistent with the thermodynamic singular point. Through our analysis of $\mu$, we find the presence of an inversion point when ${\cal H}$ is less than zero similar to the Einstein gravity and brane world scenario cases.
We determine the inversion temperature and inversion pressure for the FRW universe, and illustrate the characteristics of inversion curves and isenthalpic curves in the $T$-$P$ plane. We compared the results of isenthalpic processes with those in Einstein's theory of gravity and found a significant influence of the gravitational coupling parameter on the cooling and heating properties of the FRW universe.

It would be intriguing to explore whether the cooling and heating properties can be identified in the FRW universe filled with matter fields other than perfect fluid. It is also an interesting question that whether a negative enthalpy is a universal condition for the presence of the inversion points of the FRW universe, i.e. whether it still holds for the FRW universe in other theories of gravity or filled with other matter fields. Such investigations could yield new perspectives for testing modified gravity theories. Additionally, we are eager to determine whether cosmological observations can reveal an inversion temperature, which presents a challenging yet meaningful task. These questions will be subjects of future research. Moreover, as discussed in Appendix \ref{appA}, we have examined the constraints on the scale factor of the FRW universe derived from the inversion point. These findings hold the potential to provide valuable insights into the evolution of the universe.


\section{Acknowledgment}

We would like to thank Profs.Ya-Peng Hu, Minawar Omar and Dr.Yu-Sen An
for the useful discussions.
This work is supported by the Kashi University high-level talent research start-up fund project under Grant No. 022024002.

\appendix

\section{Constraints on the Scale Factor of the FRW Universe from Inversion Point ($\mu=0$)}\label{appA}

In this appendix, we give
the constraint on the scale factor of the FRW universe from the inversion point ($\mu=0$).

In the scenario where $\mathcal{H}<0$, it is intriguing to observe the presence of an inversion point, which corresponds to a peak in the $T$-$P$ diagram during the Joule-Thomson expansion. Exploring the deeper physical significance of this phenomenon is essential. Simplified discussions are presented below.

Equating the Hawking temperature (\ref{HT}) and the inversion temperature (\ref{invT}), we obtain
\begin{equation}
1-\dot{R}_A=-\frac{2\alpha}{R_A^2}\,,
\end{equation}
which together with (\ref{dot}) results to the following relation
\begin{equation}
1+\dot{H}R_A^2=-\frac{2\alpha}{R_A^2}\,,
\end{equation}
and by using (\ref{AH}), we finally obtain
\begin{equation}
\dot{H}+(1+2\alpha H^2)H^2=0\,, \label{B2}
\end{equation}
i.e.
\begin{equation}
\ddot{a}(t)+2\alpha \frac{\dot{a}^4(t)}{a^3(t)}=0 \,. \label{B3}
\end{equation}
It shows that the second-derivative of the scale factor $\ddot{a}(t)$ of the FRW universe is equals to $-2\alpha {\dot{a}^4(t)}/{a^3(t)}$ at the inversion point,
which means that the signs of l.h.s in Eq.(\ref{B3}) are different for $T<T_i$ and $T>T_i$.
In Einstein's theory of gravity with $\alpha=0$, the above equation (\ref{B3}) becomes $\ddot{a}=0$ at the inversion point, indicating a transition between acceleration ($\ddot{a}(t)>0$) and deceleration ($\ddot{a}(t)<0$), or vice versa \cite{Abdusattar:2021wfv}. These observations highlights the noticeable influence of gravitational coupling on the evolution of the FRW universe in a gravity with a generalized conformal scalar field.

\end{document}